\documentclass[aps,prd,showkeys,showpacs,amssymb,cite,
amsfonts,epsf,preprintnumbers,nofootinbib,superscriptaddress]{revtex4}

\usepackage[dvips]{graphicx}
\usepackage{bm,latexsym,amsmath,amssymb,amsfonts,color}

\newcommand{\be}{\begin{equation}}
\newcommand{\ee}{\end{equation}}
\newcommand{\bear}{\begin{eqnarray}}
\newcommand{\eear}{\end{eqnarray}}
\newcommand{\ba}{\begin{array}}
\newcommand{\ea}{\end{array}}

\begin{document}

\title{Static-Fluid Black Holes}

\author{Inyong Cho}
\email{iycho@seoultech.ac.kr}
\affiliation{School of Liberal Arts,
Seoul National University of Science and Technology, Seoul 01811, Korea}
\author{Hyeong-Chan Kim}
\email{hckim@ut.ac.kr}
\affiliation{School of Liberal Arts and Sciences, Korea National University of Transportation, Chungju 27469, Korea}

\begin{abstract}
We investigate black holes formed by static perfect fluid with $p=-\rho/3$.
These represent the black holes in $S_3$ and $H_3$ spatial geometries.
There are three classes of black-hole solutions,
two $S_3$ types and one $H_3$ type.
The interesting solution is the one of $S_3$ type
which possesses two singularities.
The one is at the north pole behind the horizon,
and the other is naked at the south pole.
The observers, however, are free from falling to the naked singularity.
There are also nonstatic cosmological solutions in $S_3$ and $H_3$,
and a singular static solution in $H_3$.
\end{abstract}
\maketitle

In modern cosmology, the spatial topology of the Universe is
an unresolved issue.
The recent observational data measures the curvature density as
$\Omega_k = 0.000 \pm 0.005$ ($95\%$, {\it Planck} TT+lowP+lensing+BAO)
\cite{Ade:2015xua}.
It is never manifest from the current observation
to conclude if the Universe is flat, closed, or open.
Apart from analyzing the observational data,
investigating the primordial density perturbation
in different topologies would give an insight for the topology of the Universe.
Study of inflation in the closed and the open universe
tells that there are models which is viable with the current observational data
\cite{Ellis:2002we,Ellis:2003qz,Labrana:2013oca,Bucher:1994gb,White:2014aua}.
They predict some peculiar phenomena distinguishable from flat models,
but they are still beyond the current observational resolution.
Therefore, it is not very possible to rule out any specific topology
from the cosmological studies at the current stage.

Rather than considering cosmological models,
other interesting subject would be considering a relativistic object such as a black hole
in different spatial topologies.
This might have a very significant feature
which can distinguish the spatial topology of the Universe.

The metric implying the three spatial topologies is given by
\begin{align}\label{metric0}
ds^2 = \mp dt^2 +\frac{dr^2}{1-kr^2/R_0^2} +r^2d\Omega_2^2,
\end{align}
where $k=0$ represents the flat 3-space ($R_3$),
and $k=\pm 1$ does the closed/open 3-space ($S_3/H_3$).
(The signature of $g_{tt}$ will be chosen so that we have only one time coordinate.)
This implies that the curvature of the geometry is provided by
the effective energy-momentum tensor,
\begin{align}\label{EffEM}
\bar{T}^\mu_\nu \equiv   \frac{\bar{G}^\mu_\nu}{8\pi}
= \mp \frac{1}{8\pi R_0^2} {\rm diag}(3,1,1,1),
\end{align}
where the signature $-/+$ is for $S_3/H_3$.
This implies that
\begin{align}
p=-\frac{1}{3}\rho = {\rm constant},
\end{align}
where $\rho>0$ for $S_3$ and $\rho<0$ for $H_3$.

For the closed and the open,
with the energy-momentum tensor \eqref{EffEM}
and the metric ansatz
\begin{align}\label{metricfgr}
ds^2 = -f(r)dt^2 +g(r)dr^2 +r^2d\Omega_2^2,
\end{align}
one can show that there is no static solution other than Eq.~\eqref{metric0}.
This is somewhat different from the flat case
which is a limit of the Schwarzschild solution
characterized by the mass parameter $M$.
In order to achieve a black hole configuration in the closed/open 3-space,
therefore, it is suggested to introduce a matter field.


\vspace{5pt}
In this work, we introduce a static perfect fluid
in order to see if a black hole can be formed
in the $S_3$ and $H_3$ spaces.
We introduce the spatial dependence in the energy density,
and impose a condition for the equation of state
which implies the topology of $S_3$/$H_3$, as
\begin{align}
p(r)=-\frac{1}{3}\rho(r) .
\end{align}
There are some work on perfect fluid, for example,
for charged fluid balls \cite{Bekenstein:1971ej},
for entropy of perfect-fluid objects \cite{Sorkin:1981wd,Pesci:2006sb},
for the spherical static solutions with finite-polynomial mass functions
\cite{Semiz:2008ny}, etc.

With the metric \eqref{metricfgr},
the general solutions are given by
\begin{align}
\rho(r) &= -\frac{3}{8\pi\alpha} \left\{ 1\mp \frac{2\alpha |\beta|}{r}
\left[ \beta(r^2+\alpha) \right]^{1/2} \right\}, \label{rhor}\\
f(r) &= \frac{\rho(r)}{\rho_c}, \qquad
g^{-1}(r) = -\frac{8\pi}{3} (r^2+\alpha) \rho(r). \label{fgr}
\end{align}
Note that if $\beta>0$, one needs $r^2+\alpha >0$ in Eq.~\eqref{rhor},
so $g(r)$ has the opposite signature to $\rho(r)$.
If $\beta<0$, $g(r)$ has the same signature with $\rho(r)$.
The solution for $p=-\rho/3$ in Ref.~\cite{Semiz:2008ny} is a limiting case ($\alpha\to 0$)
of Eq. \eqref{fgr}.

Let us introduce a new radial coordinate $\chi$
with which the metric is written by
\begin{align}\label{metric2}
ds^2 = -f(\chi)dt^2 +g(\chi)d\chi^2 + R_0^2b^2(\chi)d\Omega_2^2.
\end{align}
The transformation is given by $r=R_0b(\chi)$.
Depending on the values of $\alpha$ and $\beta$,
the transformed solutions of Eqs.~\eqref{rhor} and \eqref{fgr}
to the new coordinate system are classified by three types
as summarized in Tab. I.
(Here, $R_0\equiv \sqrt{|\alpha|}$ is the parameter of the 3-space curvature scale,
and $K \equiv 2R_0^2|\beta|^{3/2}$ is the positive parameter related with mass.)

\begin{table}
\begin{tabular}{|l||c|c|c|}
  \hline
                \quad Class & $\rho(\chi)$ & $f(\chi)$ & $g(\chi)$ \\ \hline\hline
  \quad $S_3$-I $\quad$
                & $\qquad \frac{3}{8\pi R_0^2} \left( 1- K \cot\chi \right) \qquad$
                & $\qquad \frac{\rho(\chi)}{\rho_c}, \quad (\rho_c>0) \qquad$
                & $\qquad \frac{3}{8\pi \rho(\chi)} \qquad$ \\ \hline
  \quad $S_3$-II
                & $\frac{3}{8\pi R_0^2} \left( 1 \mp K \tanh\chi \right)$
                & $\frac{\rho(\chi)}{\rho_c}, \quad (\rho_c<0)$
                & $-\frac{3}{8\pi \rho(\chi)}$ \\ \hline
  \quad $H_3$
                & $-\frac{3}{8\pi R_0^2} \left( 1 \mp K \coth\chi \right)$
                & $\frac{\rho(\chi)}{\rho_c}, \quad (\rho_c<0)$
                & $-\frac{3}{8\pi \rho(\chi)}$ \\
  \hline
\end{tabular}
\caption{Classification of solutions.
The signature of $\rho_c$ is chosen so that $f(\chi)$ and $g(\chi)$
 have the same signature.}
\end{table}

\begin{center} \bf{Type $S_3$-I} \end{center}

This is the case of $\alpha<0$ and $\beta<0$
($1-4\alpha^2\beta^3 >0$ always).
The transformation is performed by
\begin{equation}\label{rS3I}
r = R_0b(\chi) = R_0\sin\chi
\quad (0\leq \chi \leq \pi,\; 0\leq r \leq R_0),
\end{equation}
and the metric becomes
\begin{equation}\label{metricS3I}
ds^2 = -\frac{3}{8\pi R_0^2\rho_c} \left( 1- K \cot\chi \right) dt^2
+\frac{R_0^2}{1- K \cot\chi} d\chi^2
+R_0^2\sin^2\chi d\Omega_2^2.
\end{equation}
This is a {\bf black-hole solution} of which the horizon is located at
$\chi_h = \cot^{-1}(1/K)$.
There are two singularities;
the one is inside the horizon at $\chi=0$ where $\rho\to -\infty$,
and the other is naked at $\chi=\pi$ where $\rho\to \infty$.
The conformal diagram is plotted in Fig. 1.
Inside the horizon, $\rho<0$, and outside $\rho>0$.
The geometry near $\chi=\pi/2$ meets that of $S_3$.


\begin{center} \bf{Type $S_3$-II} \end{center}

This is the case of $\alpha<0$, $\beta>0$, and  $1-4\alpha^2\beta^3 <0$.
The transformation is performed by
\begin{equation}\label{rS3II}
r=R_0b(\chi) =R_0\cosh\chi
\quad (\chi \geq 0,\; r\geq R_0),
\end{equation}
and the metric becomes
\begin{equation}\label{metricS3II}
ds^2 = -\frac{3}{8\pi R_0^2\rho_c} \left( 1 \mp K \tanh\chi \right) dt^2
+\frac{R_0^2}{-\left( 1 \mp K \tanh\chi \right)} d\chi^2
+R_0^2\cosh^2\chi d\Omega_2^2.
\end{equation}
There are two configurations,
a black-hole solution and a nonstatic cosmological solution.

(i) {\bf Black-hole solution}:
This is the $-$ solution with $K>1$.
There exists a horizon at $\chi_h = \tanh^{-1} (1/K)$.
Inside the horizon ($0\leq \chi < \chi_h$),
$f(\chi), g(\chi) <0$ and $\rho$ is positive.
The energy density at the center is finite, $\rho(0) = 3/8\pi R_0^2$,
so there is no singularity.
The geometry near $\chi=0$ meets that of $S_3$.
Outside the horizon ($\chi> \chi_h$),
$f(\chi), g(\chi) >0$ and $\rho$ is negative and approaches a finite value.

(ii) {\bf Nonstatic cosmological solution}:
This is the $-$ solution with $K<1$, or the $+$ solution.
Everywhere, $\rho >0$ and $f(\chi), g(\chi) <0$.
Therefore, the role of $t$ and $\chi$ is exchanged,
where $\chi$ is the time coordinate now.
The radius of the Universe expands anisotropically
as $R_0\cosh\chi$ from a nonsingular state
with a finite size $R_0$.
The energy density grows from a finite value $\rho(0) = 3/8\pi R_0^2$
and approaches a positive constant.


\begin{center} \bf{Type $H_3$} \end{center}

This is the case of $\alpha>0$, $\beta>0$, and  $1-4\alpha^2\beta^3 >0$.
The transformation is performed by
\begin{equation}\label{rH3}
r= R_0b(\chi) =R_0\sinh\chi \quad (\chi \geq 0,\; r \geq 0),
\end{equation}
and the metric becomes
\begin{equation}\label{metricH3}
ds^2 = -\frac{3}{8\pi R_0^2(-\rho_c)} \left( 1 \mp K \coth\chi \right) dt^2
+\frac{R_0^2}{1 \mp K \coth\chi} d\chi^2
+R_0^2\sinh^2\chi d\Omega_2^2.
\end{equation}
There are three configurations,
a black-hole solution, a nonstatic cosmological solution,
and a singular static solution.

(i) {\bf Black-hole solution}:
This is the $-$ solution with $K<1$.
The configuration is the same as the black-hole solution in $S_3$-II,
except that the energy density blows up at the center, $\rho(0)=\infty$,
so there is a singularity.
The energy density is positive inside the horizon,
and negative outside approaching a constant where the geometry meets that of $H_3$.

(ii) {\bf Nonstatic cosmological solution}:
This is the $-$ solution with $K>1$.
The configuration is similar to that in $S_3$-II,
but the Universe expands from an initial singularity
where the energy density is infinite, $\rho=\infty$.
($\rho$ is positive all the time and approaches a constant.)
Therefore, this solution is not very interesting.

(iii) {\bf Singular static solution}:
This is the $+$ solution.
Everywhere, $f(\chi), g(\chi) >0$
and $\rho<0$.
It is singular at the center, $\rho(0)=-\infty$,
so is not interesting.


\vspace{5pt}
When $K=0$, the solutions \eqref{metricS3I}, \eqref{metricS3II} and \eqref{metricH3},
reduce to the $S_3$ and $H_3$ solutions individually,
\begin{align}
ds^2 = - dt^2 + R_0^2d\chi^2 +R_0^2 b^2(\chi) d\Omega_2^2,
\end{align}
which is the metric transformed from Eq.~\eqref{metric0}
by rescaling of the the radial coordinate $r=R_0b(\chi)$
in Eqs. \eqref{rS3I}, \eqref{rS3II} and \eqref{rH3}.
Therefore, we can conclude that the base 3-space of our solutions
are $S_3$ and $H_3$.

Since we have only one matter content,
we expect it to provide the mass in the black-hole geometry.
We analyze the near-horizon geometry, and identify $K$ with mass.
Near the horizon, $r_{\rm sch}=2M$, the Schwarzschild solution is expanded as
\begin{align}
g^{-1}(r) = 1-\frac{2M}{r} = 1-\frac{r_{\rm sch}}{r}
= \frac{1}{r_{\rm sch}}(r-r_{\rm sch}) -\frac{1}{r_{\rm sch}^2}(r-r_{\rm sch})^2 + \cdots.
\end{align}
For our fluid black-hole solution in Eqs.~\eqref{rhor} and \eqref{fgr},
the near-horizon geometry is expanded in the same way as the Schwarzschild solution
in the first order,
\begin{align}
g^{-1}(r) = \frac{1}{r_h} (r-r_h)
+ \cdots,
\qquad
r_h = \pm 2\alpha|\beta| \sqrt{\frac{\alpha\beta}{1-4\alpha^2\beta^3}},
\end{align}
where $r_h$ is the location of the horizon satisfying $\rho(r)=0$ in Eq.~\eqref{fgr},
and the $+$ sign is for $H_3$.
Identifying $r_h=r_{\rm sch}$ and using the relations $R_0 =\sqrt{|\alpha|}$
and $K = 2R_0^2|\beta|^{3/2}$, we have
\begin{align}\label{KM}
K = \left( \frac{R_0^2}{4M^2} -1 \right)^{-1/2},
\quad
\left( -\frac{R_0^2}{4M^2} +1 \right)^{-1/2},
\quad
\left( \frac{R_0^2}{4M^2} +1 \right)^{-1/2},
\end{align}
for the type $S_3$-I, $S_3$-II, and $H_3$, respectively.
Note that $M\to 0$ corresponds $K\to 0$,
which gives $S_3$ or $H_3$ geometry.

For the type $S_3$-I, the range $0<K<\infty$ corresponds to $0 <M< R_0/2$.
There is an upper limit of the mass, $M=R_0/2$.
In this limit, the horizon approaches the equator of $S_3$, $\chi_h = \cot^{-1}(1/K) \to \pi/2$.
If the mass exceeds this value, the solution of type $S_3$-I does not exist.
Instead, the type $S_3$-II is responsible for this range of mass, $M>R_0/2$.

\vspace{5pt}
Let us discuss the geodesics
of the black-hole solutions.
For simplicity, we define as
\begin{align}\label{Fchi}
F(\chi) \equiv  \frac{8\pi R_0^2}{3s}\rho(\chi),
\end{align}
where $s$ is the signature of $\rho_c$
($s=+1$ for $S_3$-I, and $s=-1$ for the rests).
The $\chi$-equation can be derived from the metric as
\begin{align}
g_{\mu\nu} \frac{dx^\mu}{d\lambda} \frac{dx^\nu}{d\lambda} = - \varepsilon,
\end{align}
where $\varepsilon = 1,0$ for timelike and null geodesics, individually.
On the $\theta=\pi/2$ plane, the $\chi$-equation becomes
\begin{align}
\frac{1}{2} \left( \frac{d\chi}{d\lambda} \right)^2 + V(\chi) = \frac{3E^2}{16\pi R_0^4|\rho_c|} \equiv \tilde{E}^2,
\qquad
V(\chi) = \frac{1}{2} F(\chi) \left[ \frac{L^2}{b^2(\chi)} +\frac{\varepsilon}{R_0^2} \right]
\end{align}
where
\begin{align}
E \equiv F(\chi) \frac{dt}{d\lambda} = {\rm constant},
\qquad
L \equiv b^2(\chi) \frac{d\phi}{d\lambda} = {\rm constant},
\end{align}
defined from the $t$- and $\phi$-equations.
We summarize the effective potential $V(\chi)$ in Tab. II.

\begin{table}
\begin{tabular}{|l||c|c|}
  \hline
                \quad Class & $F(\chi)$ & $V(\chi)$ \\ \hline\hline
  \quad $S_3$-I $\quad$
                & $\qquad 1- K \cot\chi \qquad$
                & $\qquad \frac{1}{2} (1-K\cot\chi) \left( \frac{L^2}{\sin^2\chi} +\frac{\varepsilon}{R_0^2} \right) \qquad$ \\ \hline
  \quad $S_3$-II
                & $-1 + K \tanh\chi$
                & $\frac{1}{2} (-1+K\tanh\chi) \left( \frac{L^2}{\cosh^2\chi} +\frac{\varepsilon}{R_0^2} \right)$ \\              \hline
  \quad $H_3$
                & $1 - K \coth\chi$
                & $\frac{1}{2} (1-K\coth\chi) \left( \frac{L^2}{\sinh^2\chi} +\frac{\varepsilon}{R_0^2} \right)$ \\
  \hline
\end{tabular}
\caption{Effective potential $V(\chi)$}
\end{table}

\begin{center} \bf{Type $S_3$-I} \end{center}

For $L=0$, the potential $V$ for timelike geodesics monotonically grows
from $-\infty$ (north pole) to $\infty$ (south pole).
All the paths fall into the center of the black hole.
(See Fig. 2.)

For $L\neq 0$, both of the timelike and null geodesics have a similar potential shape.
If $K \geq 1/\sqrt{3[\varepsilon/(L^2R_0^2)+1]}$,
the potential is similar to that of the timelike geodesics for $L=0$.
All the paths fall into the center of the black hole.
If $K < 1/\sqrt{3[\varepsilon/(L^2R_0^2)+1}]$, however,
there is a potential well outside the horizon $\chi_h$.
Therefore, a stable orbit around the black hole is possible.

The outgoing radial null rays ($L=0$) can reach
the naked singularity at the south pole ($\chi =\pi$).
However, none of the other geodesics can reach that singularity.
The observer can only see the light signal from the singularity at the south pole.

As the matter around the black hole falls into the horizon,
the mass parameter $K$ increases as in Eq.~\eqref{KM}.
If the value of $K$ exceeds the above critical value $1/\sqrt{3\varepsilon/(L^2R_0^2)+1}$,
all the matter around the black hole is destined to fall into the black hole except the radial outgoing null rays.
The horizon size gets larger until it reaches the equator, $\chi_h = \cot^{-1}(1/K) \to \pi/2$,
as the mass reaches the upper bound, $K\to \infty$ ($M\to R_0/2$).
After that, this type of black-hole solution does not count.

\begin{center} \bf{Type $S_3$-II and Type $H_3$} \end{center}

For $S_3$-II and $H_3$, their geodesic motions are similar.
For the timelike geodesics, if $L^2 \leq K/(2|K-1|R_0^2)$,
the potential $V$ monotonically grows
from $V(0)=-(L^2+1/R_0^2)/2$ for $S_3$-II
and $V(0)=-\infty$ for $H_3$,
and approaches a constant $V(\infty) = |K-1|/(2R_0^2)$.
(Note that $K>1$ for the $S_3$-II black hole,
and $0<K<1$ for the $H_3$ black hole.)
The ingoing geodesics fall into the black hole,
while the outgoing geodesics bounce back or escape to infinity
depending on their energy level $\tilde{E}$.
(See Fig. 2.)

If $L^2 > K/(2|K-1|R_0^2)$,
there exists a potential bump outside the horizon.
The potential for the null geodesics ($L\neq 0$) is also similar to this case
but with different boundary values, $V(0)=-L^2/2$ and $V(\infty)=0$.
Unlike the type $S_3$-I,
there is no potential well, so the stable orbit is absent in these two types.
The ingoing geodesics can bounce back to infinity
if $\tilde{E}$ is not large enough.

\vspace{5pt}
Let us study the stability of the black holes.
We introduce linear spherical scalar perturbations as following.
The metric is written by
\begin{align}
ds^2 = -f(t,\chi)dt^2 + g(t,\chi) d\chi^2 + R_0^2 b^2(\chi) d\Omega_2^2.
\end{align}
The perturbations for the metric are introduced as
\begin{align}
f(t,\chi) &= f_0(\chi) + \epsilon f_1(t,\chi), \label{p1}\\
g(t,\chi) &= R_0^2 \big[ g_0(\chi) + \epsilon g_1(t,\chi) \big], \label{p2}
\end{align}
where $\epsilon$ is a small parameter,
and the subscript $0$ stands for the background solutions
obtained earlier.
Using the definition in Eq.~\eqref{Fchi}
for $F(\chi) = 8\pi R_0^2\rho_0(\chi)/3s$,
where $\rho_0(\chi)$ is the background solution in Tab. I,
we have
\begin{align}
f_0(\chi) &= \frac{\rho_0(\chi)}{\rho_c} = \frac{3s}{8\pi R_0^2\rho_c}F(\chi), \\
g_0(\chi) &= \frac{1}{F(\chi)}.
\end{align}
The energy-momentum tensor is given by
\begin{align}
T^{\mu\nu} = (\rho +p)u^\mu u^\nu + pg^{\mu\nu},
\end{align}
where the velocity four-vector is given by
\begin{align}
u^\mu = \big[ u^0(t,\chi),u^1(t,\chi),0,0 \big].
\end{align}
For the barotropic fluid, $p=w\rho$,
the perturbations for the energy density and the four-velocity are given by
\begin{align}
\rho(t,\chi) &= \rho_0(\chi) +\epsilon \rho_1(t,\chi), \label{p3}\\
u^0(t,\chi) &= u_0^0(\chi) +\epsilon u_1^0(t,\chi), \label{p4}\\
u^1(t,\chi) &= u_0^1(\chi) +\epsilon u_1^1(t,\chi). \label{p5}
\end{align}
For the comoving background fluid, we have $u_0^1(\chi)=0$.
From the normalization $u^\mu u_\mu=-1$, we have
$u_0^0(\chi)=1/\sqrt{f_0(\chi)}$ and
$u_1^0(t,\chi)=-f_1u_0^0/(2f_0) = -f_1/(2f_0^{3/2})$.

Now we apply the perturbations \eqref{p1}, \eqref{p2}, and \eqref{p3}-\eqref{p5},
and expand the field equations in the first order of $\epsilon$.
From the $(0,1)$ component of the Einstein's equation, one gets
\begin{align}
u_1^1(t,\chi) = -\sqrt{\frac{2\pi R_0^2\rho_c}{3}} \frac{\dot{g_1}b'F}{s^2b\sqrt{F}}.
\end{align}
Therefore, the perturbations of the four-vector, $u_1^0$ and $u_1^1$, in Eqs.~\eqref{p4} and \eqref{p5}
are completely described by the metric perturbations and the background functions.
The components of the conservation equation $\nabla_\mu T^{\mu\nu} =0$
are equivalent to the Einstein equations,
so one may simply solve the Einstein equations.
After manipulating the Einstein equations, one finally gets
the single perturbation equation.
We introduce the perturbation in the form,
\begin{align}
g_1(t,\chi) = e^{i\omega t} \varphi(\chi),
\end{align}
and transform the radial coordinate and the amplitude function as
\begin{align}
z = \int^\chi_0 \frac{d\chi}{2F(\chi)}, \qquad
\Phi(z) = N \frac{F(\chi)b'(\chi)}{z} \varphi(\chi),
\end{align}
where $N$ is a normalization constant.
Then we get the perturbation equation from the Einstein's equation
in the nonrelativistic Schr\"odinger-type,
\begin{align}\label{PE2}
\left[ -\frac{1}{2}\frac{d^2}{dz^2} - \frac{1}{z}\frac{d}{dz}
+U(z) \right] \Phi(z) = -\frac{\omega^2}{\sigma} \Phi(z)
=-8\pi R_0^4|\rho_c| \omega^2 \Phi(z) \equiv \Omega \Phi(z),
\end{align}
where $\sigma \equiv 1/(8\pi R_0^4\rho_c s) = 1/(8\pi R_0^4|\rho_c|) >0$,
and the potential is
\begin{align}
U[z(\chi)] = F^2 \left[ -\frac{F''}{F} +\left( \frac{F'}{F} \right)^2
-\frac{F'}{F} \left( 2\frac{b''}{b'} -\frac{b'}{b} \right) -\frac{b'''}{b'}
+2 \left( \frac{b''}{b'} \right)^2 -2\frac{b''}{b} +2\left( \frac{b'}{b} \right)^2
\right].
\end{align}
Since there always exists a positive eigenvalue $\Omega$ for any type of potential $U$,
i.e., $\omega^2 <0$,
this system is {\it unconditionally unstable}.

The stability analysis adopted here tells only the existence of the instability.
It does not give any information in which direction the system will evolve
due to the instability.
One may guess that the instability is from the nature of perfect fluid.
When perfect fluid is introduced in the Universe,
it usually drives the expansion of the Universe.
Although there exists a static solution with perfect fluid as ours,
a small perturbation may invoke fluid to drive the spacetime to expand.
This could be the source of our instability.

In order to catch such a nature of instability,
other methods of stability check need to be employed.
One complete method will be the numerical investigation.
Other than that, we can consider a bit simpler way
to check the direction of the instability in the linear order.
In the stability investigation, we introduce a bit different
metric perturbations which admit the evolution of the background universe
as well as the black-hole size.
The first one is done by introducing a scale factor $S(t)$,
and the second is done by letting the mass scale $K$ be time-dependent, as
\begin{align}
S^2(t) \equiv 1+ \epsilon a(t), \qquad
K \to K+\epsilon\kappa(t).
\end{align}
Then the perturbed metric can be written as
\begin{align}
ds^2 = -f_0\big[\chi,\kappa(t)\big]dt^2
+ S^2(t) \left\{ g_0\big[\chi,\kappa(t)\big] d\chi^2 + R_0^2 b^2(\chi) d\Omega_2^2 \right\}.
\end{align}
The solutions to the Einstein equations are given by
\begin{align}
a(t) = a_1 t + a_0, \qquad
\kappa(t) = \kappa_1 t + \kappa_0.
\end{align}
Both of the scale factor and the mass parameter evolve linearly in time.
According to $a(t)$, the background universe may expand, or shrink.
According to $\kappa(t)$, the coordinate of the horizon location $\chi_h$
may increase, decrease, or remain constant.
Assuming the background universe expands due to the instability,
the black hole may change its physical size in both directions,
or may remain unchanged.
Beyond the linear order, we still need further investigation
via, for example, numerical study.

\vspace{5pt}
In this letter, we investigated the black-hole solutions
in a closed/open 3-space.
It is formed by introducing a static perfect fluid
with the equation of state $p(r)=-\rho(r)/3$.
There are three classes of solutions that we named as $S_3$-I, $S_3$-II, and $H_3$.
All the solutions are characterized mainly by a parameter $K$ which is related with mass.
When the mass parameter $K$ is turned off,
the remaining geometry possesses only the 3-space curvature scaled by the parameter $R_0$.
The solutions are classified by this curvature scale into three types, $S_3$-I, $S_3$-II, and $H_3$.
The first two types represent the closed 3-space,
and the last type does the open one.

In all classes, there is a black-hole solution.
The interesting one is the $S_3$-I type.
There are two singularities at both poles.
The one is hidden by the horizon, and the other is naked.
However, except the radial null geodesic with zero angular momentum ($L=0$),
the naked one is not accessible by timelike and null geodesics.
They are destined to fall into the black hole,
or may have a stable orbit with a large angular momentum.
The black-hole of the $S_3$-II type has no singularity,
but the energy density is negative outside the horizon.
The black hole of the $H_3$ type has a singularity at the center.

For the $S_3$-I type black hole,
all the energy conditions are satisfied outside the horizon.
The existence of the naked singularity at the south pole
looks violating the singularity theorem.
This must be due to that the background 3-space is $S_3$.
Although $\rho<0$ inside the horizon,
the energy density is positive there because $p_1=p$ is regarded
as the relevant energy density in this nonstatic region.

Other than the black-hole solution,
there is a nonstatic cosmological solution of $S_3$-II and $H_3$ types,
which represents an expanding universe.
The interesting one is the $S_3$-II type,
which expands from a nonsingular finite size of the Universe,
while there is an initial singularity at the $H_3$ type.

Our black holes are unconditionally unstable.
We suspect that the instability is from the nature of perfect fluid
which usually drives the background spacetime to expand.
Our stability study shows some hints of this kind of evolution.
The background universe expands due to the instability
while the physical size of the black hole
may either change, or remain fixed
depending on the conditions imposed.
The full stability study requires numerical investigations.

Our work was focused only on a single matter field which can be
responsible both for the 3-space topology and the black-hole curvature.
This was done by imposing $p(r)=-\rho(r)/3$ on fluid.
The black holes in this work are a bit far from providing enough phenomenological information
for discussing the topology of the currently observed universe.
However, continuing study may give some insights about it.
For example, one can introduce the matter fields separately, i.e.,
the one for the 3-space topology with a constant energy density,
$\mathfrak{p}=-\varrho/3={\rm constant}$,
and the other for the black-hole curvature $p(r)=w\rho(r)$
with an arbitrary value of $w$.
Also, one can introduce a scalar field rather than fluid.
Then the black holes may have different geometry and stability.
We will get back to these soon in the future.

\acknowledgements
We are grateful to Gungwon Kang for useful discussions.

\begin{figure*}[btph]
\begin{center}
\includegraphics[width=0.8\textwidth]{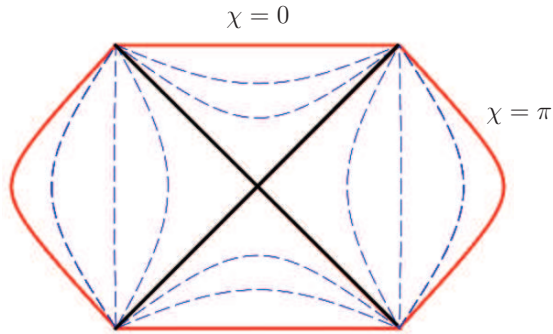}
\end{center}
\vspace{-5.9in}
\caption{Conformal diagram for $S_3$-I black hole.
There are two singularities at $\chi=0$ (north pole: center of the black hole)
and at $\chi=\pi$ (south pole: naked).
The solid diagonal lines represent the horizon at $\chi_h=\cot^{-1}(1/K)$.
The dashed lines are the $\chi$-constant lines.
The $t$-constant lines are straight lines passing the center of the diagram (not shown).
Outside the horizon, except the outgoing null rays,
none of the geodesics can reach the naked singularity at the south pole.
}
\end{figure*}
\begin{figure*}[btph]
\begin{center}
\includegraphics[width=.4\textwidth]{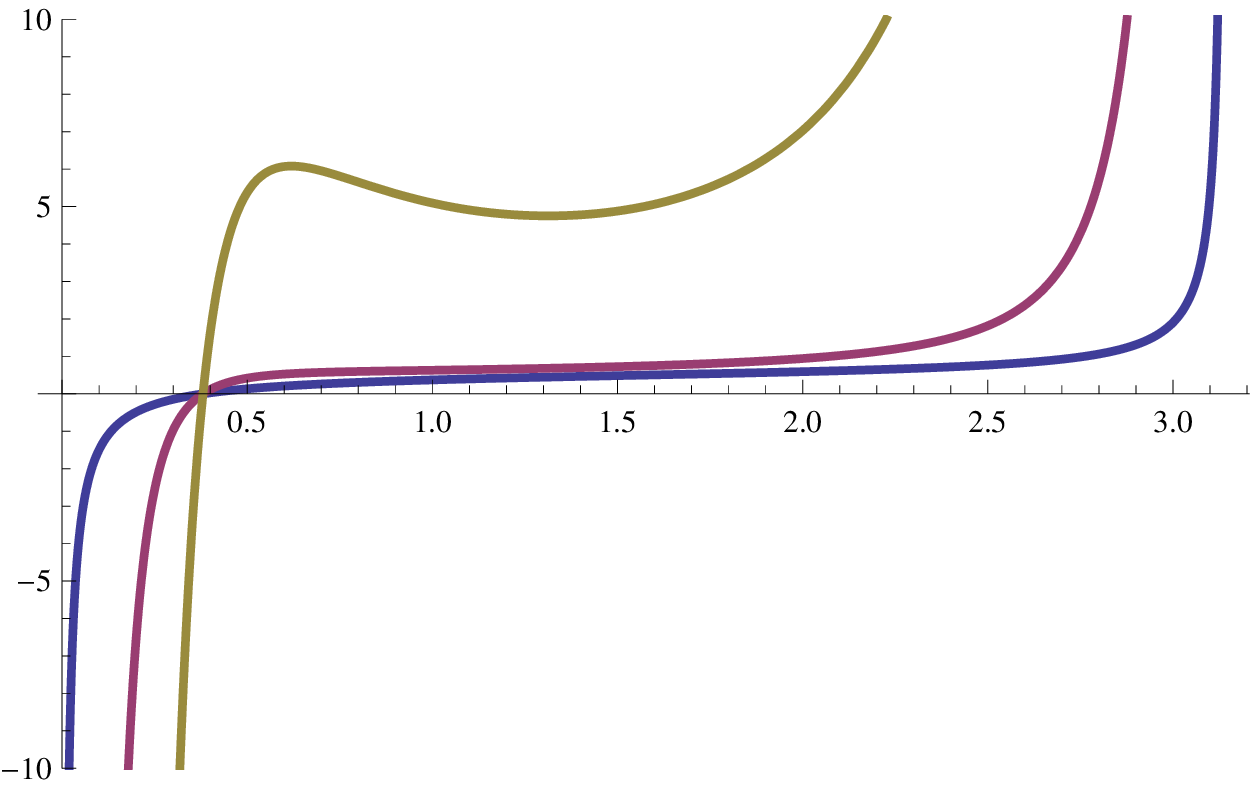}
\hspace{1cm}
\includegraphics[width=.4\textwidth]{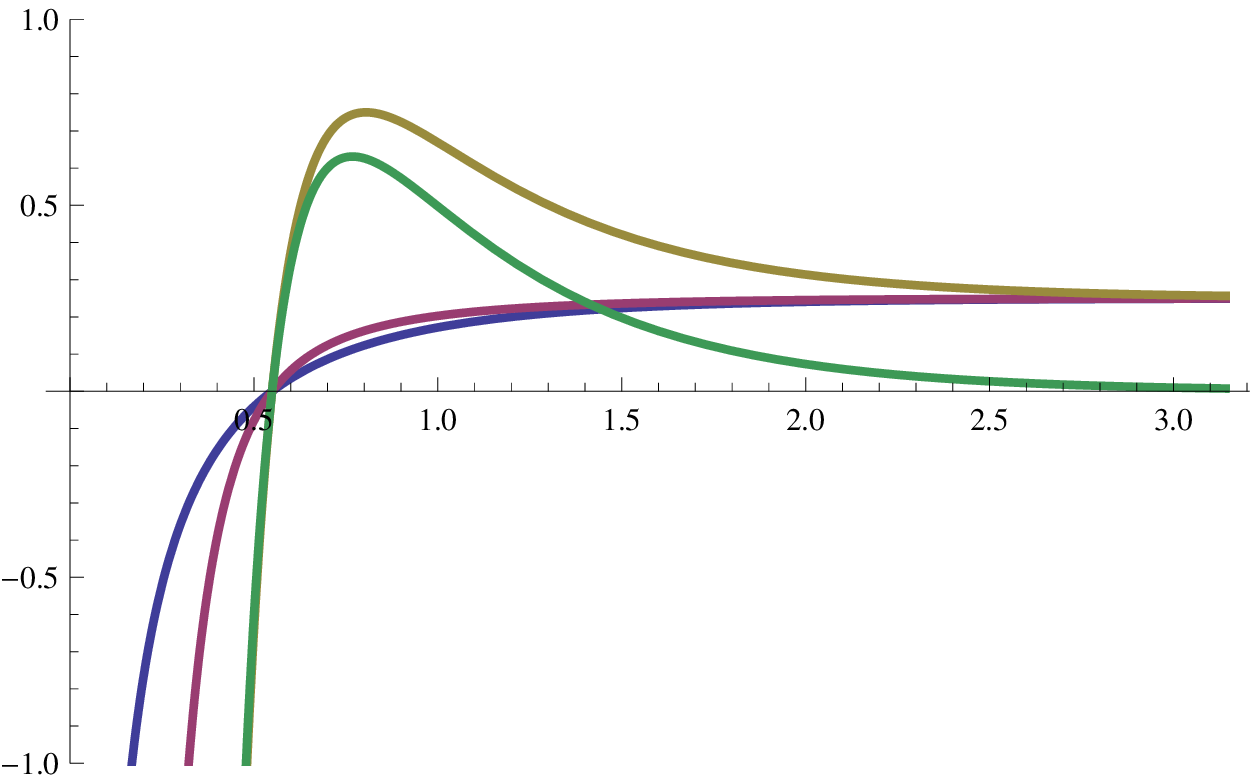}
\end{center}
\caption{Plot of the effective potential $V(\chi)$.
(a) $S_3$-I (left panel): timelike geodesics for $R_0=1$ and $K=0.4$:
from the bottom, $L=0$, $0.7$, $3$.
For the null geodesics ($L\neq 0$),
the potential shapes are similar.
(b) $H_3$ (right panel): timelike and null geodesics for $R_0=1$ and $K=0.5$.
For timelike, $L=0$, $0.5$, $2$ from the bottom,
and for null, $L=2$ (green).
For $S_3$-II, the potential $V$ is very similar,
but with a finite central value, $V(0)=-(L^2+\epsilon/R_0^2)/2$.
}
\end{figure*}


\begin{thebibliography}{99}

\bibitem{Ade:2015xua}
  P.~A.~R.~Ade {\it et al.} [Planck Collaboration],
  Astron.\ Astrophys.\  {\bf 594}, A13 (2016)
  doi:10.1051/0004-6361/201525830
  [arXiv:1502.01589 [astro-ph.CO]].

\bibitem{Ellis:2002we}
  G.~F.~R.~Ellis and R.~Maartens,
  Class.\ Quant.\ Grav.\  {\bf 21}, 223 (2004)
  doi:10.1088/0264-9381/21/1/015
  [gr-qc/0211082].

\bibitem{Ellis:2003qz}
  G.~F.~R.~Ellis, J.~Murugan and C.~G.~Tsagas,
  Class.\ Quant.\ Grav.\  {\bf 21}, no. 1, 233 (2004)
  doi:10.1088/0264-9381/21/1/016
  [gr-qc/0307112].

\bibitem{Labrana:2013oca}
  P.~Labrana,
  Phys.\ Rev.\ D {\bf 91}, no. 8, 083534 (2015)
  doi:10.1103/PhysRevD.91.083534
  [arXiv:1312.6877 [astro-ph.CO]].

\bibitem{Bucher:1994gb}
  M.~Bucher, A.~S.~Goldhaber and N.~Turok,
  Phys.\ Rev.\ D {\bf 52}, 3314 (1995)
  doi:10.1103/PhysRevD.52.3314
  [hep-ph/9411206].


\bibitem{White:2014aua}
  J.~White, Y.~l.~Zhang and M.~Sasaki,
  Phys.\ Rev.\ D {\bf 90}, no. 8, 083517 (2014)
  doi:10.1103/PhysRevD.90.083517
  [arXiv:1407.5816 [astro-ph.CO]].

\bibitem{Bekenstein:1971ej}
  J.~D.~Bekenstein,
  Phys.\ Rev.\ D {\bf 4}, 2185 (1971).
  doi:10.1103/PhysRevD.4.2185

\bibitem{Sorkin:1981wd}
  R.~D.~Sorkin, R.~M.~Wald and Z.~J.~Zhang,
  Gen.\ Rel.\ Grav.\  {\bf 13}, 1127 (1981).
  doi:10.1007/BF00759862

\bibitem{Pesci:2006sb}
  A.~Pesci,
  Class.\ Quant.\ Grav.\  {\bf 24}, 2283 (2007)
  doi:10.1088/0264-9381/24/9/009
  [gr-qc/0611103].

\bibitem{Semiz:2008ny}
  I.~Semiz,
  Rev.\ Math.\ Phys.\  {\bf 23}, 865 (2011)
  doi:10.1142/S0129055X1100445X
  [arXiv:0810.0634 [gr-qc]].




\end{thebibliography}
\end{document}